\documentstyle[amstex,amssymb,11pt,epsfig]{article}

\newtheorem{pro}{Proposition}

\newcommand{\dif}{\operatorname{d}}

\newcommand{\diag}{\operatorname{diag}}
\newcommand{\sgn}{\operatorname{sgn}}

\newcommand{\bt}{\boldsymbol t}
\newcommand{\bx}{\boldsymbol x}
\newcommand{\bX}{\boldsymbol X}
\newcommand{\bn}{\boldsymbol n}

\newcommand{\bell}{{\boldsymbol\ell}}
\newcommand{\be}{{\boldsymbol e}}

\newcommand{\bOmega}{{\boldsymbol \Omega}}

\newcommand{\bpsi}{{\boldsymbol \psi}}
\newcommand{\bPsi}{{\boldsymbol \Psi}}
\newcommand{\blambda}{{\boldsymbol \lambda}}
\newcommand{\cL}{{\cal L}}
\newcommand{\cF}{{\cal F}}
\newcommand{\C}{{\Bbb C}}
\newcommand{\R}{{\Bbb R}}
\newcommand{\Z}{{\Bbb Z}}

\newcommand{\id}{\operatorname{id}}

\newcommand{\Sgot}{{\cal S}}

\newcommand{\gT}{{\Bbb V}}
\newcommand{\gX}{{\Bbb X}}

\newcommand{\D}{{\partial}}
\newcommand{\Cninf}{\C^{N\raisebox{.4mm}{\scriptsize$\cdot\infty$}}}

\begin{document}

\title{Charged Free Fermions,
Vertex Operators   and\\ Classical  Theory of Conjugate Nets}
\author{Adam Doliwa$^{1,2}$, Manuel Ma\~nas$^{3,4}$,
Luis Mart\'\i nez Alonso$^4$,\\ Elena Medina$^5$ and Paolo Maria
Santini$^{1,6}$ \\ \\ $^1$Istituto Nazionale di Fisica Nucleare,
Sezione di Roma,\\ P-le. Aldo Moro 2, I-00185, Italy\\ $^2$Instytut
Fizyki Teoretycznej, Uniwersytet Warszawski\\ ul. Ho\.{z}a 69,
00-681 Warszawa, Poland\\ $^3$Departamento de Matem\'{a}tica Aplicada y
Estad\'{\i}stica\\ E. U. I. T. Aerona\'{u}tica, Universidad Polit\'{e}cnica de
Madrid\\ E28040-Madrid, Spain\\
 $^4$Departamento de F\'\i sica
Te\'orica, Universidad Complutense\\ E28040-Madrid, Spain\\
$^5$Departamento de Matem\'aticas, Universidad de C\'{a}diz\\
E11510-C\'{a}diz, Spain\\ $^6$Dipartimento di Fisica, Universit\`a di
Catania \\ Corso Italia 57, I-95129 Catania, Italy}


\date{PACS \#: 02.40.+m; 03.70.+k \\
{\bf Keywords}: vertex operators, conjugate nets
}
\maketitle

\begin{abstract}
We show that the quantum field theoretical formulation of the
$\tau$-function theory
has a geometrical interpretation within the classical
transformation theory of conjugate nets. In particular, we prove
that i)~the partial charge transformations  preserving the neutral
sector are Laplace transformations, ii)~the  basic vertex operators
are L\'{e}vy and adjoint L\'{e}vy transformations and iii)~ the diagonal
soliton vertex operators generate fundamental transformations.
 We also show that the bilinear identity for the multicomponent
Kadomtsev-Petviashvili hierarchy becomes, through a generalized
Miwa map,  a bilinear identity for the multidimensional
quadrilateral lattice equations.
\end{abstract}
\newpage%
\section{Introduction}
The notion of $\tau$-function  is a cornerstone in the theory of
integrable systems since its discovery in 1976 by Hirota
\cite{hirota-tau}. It allows us to reformulate many integrable
equations as bilinear equations for $\tau$-functions and provides
suitable methods for finding soliton solutions. In a series of
papers \cite{monodromy}, Sato, Jimbo and Miwa introduced the
concept of $\tau$-function in the framework of quantum field theory
of free fermions in two-dimensional space-time. They were motivated
by the fact that certain limits of correlation functions of the
two-dimensional Ising model provide solutions of the Painlev\'{e} III
equation \cite{ising}. As a result, they revealed an unexpected
link with the isomonodromonic deformation theory of linear
differential equations, a classical mathematical subject started by
Riemann in the last century and investigated by Schlessinger
\cite{schl}, Fuchs \cite{fuchs} and Garnier \cite{garnier} among
others.
 Later on \cite{kp}, it was shown  that the same quantum field
 theoretical
formulation applies equally well for describing  the
Kadomtsev-Petviashvili (KP) hierarchy and its multicomponent
extension  in terms of $b$-$c$ systems of ghosts fields. This
description turned out to be related with certain aspects of string
theory as, for example, the connection between  bosonic string
amplitudes and Hirota's difference equation \cite{saito,m,h}.
 Furthermore, the
Grassmannian model \cite{sw} of the $\tau$-function theory is
strongly related with the operator formalism in string theory
and conformal field theory 
\cite{w,ag}. In this context, it is worth mentioning that the
Korteweg-de Vries hierachy (a reduction of KP) has appeared in a
non perturbative description of two-dimensional quantum gravity
\cite{2dqg}.

On the other hand, the theory of conjugate nets
\cite{Darboux,bianchi,Eisenhart} is a classical subject in
differential geometry  developed by distinguished geometers of the
last century and the begining of the present one (Gauss, Lam\'{e},
Bonet, Cayley, Bianchi, Darboux, Eisenhart \dots). In particular,
the transformation theory of conjugate nets is  a well established
subject \cite{Eisenhart}. The orthogonal reduction is strongly
connected with the theory of quasilinear systems of hydrodynamic
type, which in the two-dimensional case includes the Euler
equations of an incompressible fluid. Riemann devoted part of his
work to the classification of these systems in terms of Riemann
invariants, this was extended to the multidimensional case by
Tsarev \cite{tsarev} showing that the classification problem of
Hamiltonian systems of hydrodynamic type is equivalent to the study
of orthogonal nets. Let us also mention that conjugate nets are
connected with the description of the three wave resonant
interaction. The discrete analogues of the conjugate nets,
quadrilateral lattices, are central objects of the integrable
discrete geometry which is developing nowadays \cite{DQL1,TQL}.
Finally,  Egoroff systems (a particular type of conjugate nets)
have recently found application  in topological field theory;
namely, in the resolution of indescomposable
Witten-Dickgraff-Verlinde-Verlinde associativity equations
\cite{dubrovin}.

The aim of this paper  is to show that the basic operations
associated with the multicomponent KP theory have a distinguished
geometric interpretation in terms of conjugate nets, their
transformations and discretisations. Thus, the quantum field
theoretical scheme introduced by the Kyoto school is strongly tyed
up not only with monodromy problems and integrable systems but also
with the developments that started last century in the arena of
classical differential geometry.

The layout of the paper is as follows,  the first two sections have
a introductory character: in \S 2, we introduce  standard material
on the quantum field theory description of the multicomponent KP
hierarchy (for an alternative approach see \cite{kv,bk})  and, in
\S 3, the theory of conjugate nets and its transformations, namely
Laplace, L\'{e}vy, adjoint L\'{e}vy and fundamental transformations. The
next two sections, \S 4 and \S 5, contains the main results of our
paper. In
\S 4 we show that:
\begin{enumerate}
\item  The partial charge transformations preserving total charge, i. e.
the Schlessinger transformations, are Laplace transformations.
\item  The basic vertex operators can be identified with the
L\'{e}vy and adjoint L\'{e}vy transformations.
\item The diagonal soliton vertex operator generates a fundamental
transformation.
\end{enumerate}
In this manner a complete list of  equivalences among  the basic
transformations of both schemes arises. We underline that these
correspondences are linked with a series of Fay identities for the
$\tau$ function.

Finally, in \S 5, we extend the Miwa transformation to the
multicomponent case to obtain the quadrilateral lattice. Having
this result, as we shall exhibit,  a natural geometrical
interpretation.

\section{$b$-$c$ systems and $\tau$ functions}
\label{sec:mKP}
The $b$-$c$ system of quantum fields, which appears as the system
of ghost fields in string theory, is constructed in terms of the
anticommutation relations
\begin{align*}
&\{b_i(z),c_j(z^\prime)\}=\delta_{ij}\delta(z-z^\prime),\\
&\{b_i(z),b_j(z^\prime)\}=\{c_i(z),c_j(z^\prime)\}=0,
\end{align*}
where $b_i(z)$ and $c_i(z)$, $i=1,\dotsc,N$, are free charged
fermion fields defined on the unit circle $S^1$, and
$\delta(z-z^\prime)$ is the Dirac distribution on $S^1$.

The Clifford algebra generated by the $b$-$c$ system admits a
representation in terms of bosonic variables. In this
representation the fields act on the Fock space $\cal F$ of
complex-valued functions
\[
\tau=\sum_\bell\tau(\bell,\bt)\blambda^\bell,
\]
with
\begin{gather*}
\bell:=(\ell_1,\dotsc,\ell_N)\in\Z^N,\\
\bt:=(\bt_1,\dotsc,\bt_N)\in\C^{N\raisebox{.4mm}{
\scriptsize$\cdot\infty$}},\quad
\bt_i:=(t_{i,1},t_{i,2},\dotsc)\in\C^\infty,\\
\blambda:=(\lambda_1,\dots,\lambda_N)\in\C^N,\quad
\blambda^\bell:=
\lambda_1^{\ell_1}\dotsb\lambda_N^{\ell_N}
\end{gather*}

 The representation of the $b$-$c$ generators takes the form
  \cite{djkm}:
\[
b_i(z):=X_i(z)S_i(z)\prod_{j>i}P_j,\quad
c_i(z):=X^*_i(z)S^*_i(z)\prod_{j>i}P_j,\quad i=1,\dots,N,
\]
where
\begin{gather*}
X_i(z):=\exp(\xi(x,\bt_i))\gT_i^-(z),\quad
X^*_i(z):=\exp(-\xi(x,\bt_i))\gT_i^+(z),\\ S_i(z):=\lambda_i
z^{\lambda_i\D/\D\lambda_i},\quad S^*_i(z):=\frac{1}{\lambda_i}
z^{1-\lambda_i\D/\D\lambda_i},\\
P_i(\blambda^\bell)=(-1)^{\ell_i}\blambda^\bell.
\end{gather*}
Here we are denoting
\[
\xi(z,\bt_i):=\sum_{n=1}^\infty z^n t_{i,n},
\]
and $\gT_i^\pm$ are operators defined by
\[
\gT_i^{\pm}(z)f(\bt):=f(\bt\pm[1/z]\be_i),\quad
\left[ 1/z \right] := \left( \frac{1}{z}, \frac{1}{2z^2}, \frac{1}{3z^3},
\ldots \right),
\]
being  $\{\be_i\}_{i=1}^N$  the canonical generators of $\C^N$.

Alternatively, the action of the $b$-$c$ system can be formulated
as
\begin{align*}
b_i(z)\tau(\bell,\bt)&:=(-1)^{\sum_{j>i}\ell_j} z^{\ell_i-1}
\exp(\xi(z,\bt_i))
\tau(\bell-\be_i,\bt-\left[ 1/z \right]\be_i),\\
c_i(z)\tau(\bell,\bt)&:=(-1)^{\sum_{j>i}\ell_j}
z^{-\ell_i}\exp(-\xi(z,\bt_i))
\tau(\bell+\be_i,\bt+\left[ 1/z \right]\be_i).
\end{align*}

The Fock space decomposes into a direct sum of charge sectors
\[
{\cal F}=\bigoplus_{q\in\Z}{\cal F}_q, \quad {\cal F}_q=\{\tau\in
{\cal F}: Q\tau=q\tau\},
\]
where the total charge operator $Q:=\sum_{i=1}^NQ_i$ is the sum of
$N$ commuting partial charges $Q_i=\lambda_i\D/\D\lambda_i$,
$i=1,\dots, N$; they correspond to the  $N$ different flavours of
fermions of the model.

The $N$-component KP hierarchy can be formulated as the following
bilinear identity
\begin{equation}\label{bilineal}
{\cal B}(\tau\otimes\tau)=0,\quad \tau\in{\cal F}_0,
\end{equation}
where
\[
{\cal B}:=\int_{S^1}\frac{\dif z}{z}\sum_{i=1}^N b_i(z)\otimes
c_i(z).
\]
In terms of the components $\tau(\bell,\bt)$ we have
\begin{multline}\label{bi1}
\int_{S^1}\dif z\sum_{m=1}^N(-1)^{\sum_{j>m}\ell_j+\ell_j^\prime}
\exp(\xi(z,\bt_m-\bt_m^\prime))z^{\ell_m-\ell_m^\prime-2}\\
\times\tau(\bell-\be_m,\bt-\left[ 1/z \right]\be_m)
\tau(\bell^\prime+\be_m,\bt^\prime+\left[ 1/z \right]\be_m)=0,
\end{multline}
for any $\bt$, $\bt'$ and $\bell$, $\bell'$ such that
$\ell_1+\dots+\ell_N-1=\ell'_1+\dots+\ell'_N+1=0$.

Let us mention that, as it  is well known,
 in the fermionic picture
of the $\tau$-function one can define a vacuum  in such a manner
that the $\tau$-function  becomes the vacuum expectation value of a
suitable  element of the Clifford algebra, that evolves
 according to the time generator defined by certain
  Hamiltonian ${\cal H}(\bt)$ \cite{djkm}.

It can be shown that
\[[X_i\otimes X_i,{\cal B}]=
[X_i^*\otimes X_i^*,{\cal B}]=0;
\]
i. e.,  for any solution $\tau$
 of (\ref{bilineal}) the functions  $X_i\tau$
and $X_i^*\tau$ satisfy (\ref{bilineal}) as well. Hence, the vertex
operators $X_i$ and $X_i^*$, $i=1,\dots, N$, constitute a set of
symmetries of (\ref{bilineal}).

In our subsequent analysis we will fix a given $\bell$ and denote
\begin{equation*} \label{l-fix}
\begin{aligned}
\tau(\bt) & :=\tau(\bell,\bt), \\
\tau_{ij}(\bt) &:= \Sgot_{ij}\tau(\bt)
 := \tau(\bell+\be_i-\be_j,\bt).
\end{aligned}
\end{equation*}

Observe that the vectors $\alpha_{ij} =\be_i-\be_j$ are the roots
of the $A_{N-1}$ root system, so that any linear combination of
them with integer coefficients is a point in the corresponding root
lattice. The shift operators $\Sgot_{ij}$ along the root lattice
vectors $\alpha_{ij}$ correspond to the so called Schlessinger
transformations in monodromy theory~\cite{djkm1,jm,JvL} and satisfy
the following relations
\begin{align}
\Sgot_{ij}\circ\Sgot_{ji} & = \id \label{ij-ji} \\
\Sgot_{ij}\circ\Sgot_{jk} = \Sgot_{ik}, \; \; & \; \;
\Sgot_{ij}\circ\Sgot_{ki} = \Sgot_{kj}. \label{ijk}
\end{align}
This root lattice models all the possible transformations in the
partial charges that do not alter the neutral character of the
asembly of fermions. Moreover, ${\cal S}_{ij}$ are obvious
symmetries of (\ref{bilineal}).

The $N\times N$ matrix Baker function $\psi$ and its adjoint
$\psi^*$ can be defined in terms of the $\tau$ function as
\begin{align}\label{baker1}
\psi_{ij}(z,\bt)&=\varepsilon_{ij}z^{\delta_{ij}-1}
\frac{\tau_{{ij}}(\bt-\left[ 1/z \right]\be_j)}{\tau(\bt)}\exp(\xi(z,\bt_j))
,\\ \notag
\psi_{ij}^*(z,\bt)&=\varepsilon_{ji}z^{\delta_{ij}-1}\exp(-\xi(z,\bt_i))
\frac{\tau_{{ij}}(\bt+\left[ 1/z \right]\be_i)}{\tau(\bt)},
\end{align}
where $\varepsilon_{ij}:=\sgn(j-i)$, $j\neq i$
($\varepsilon_{ii}:=1$).

Observe that we have
\begin{equation}\label{baker}
\begin{aligned}
\psi(z,\bt)&:=\chi(z,\bt)\psi_0(z,\bt),\\
\psi^*(z,\bt)&:=\psi_0(z,\bt)^{-1}\chi^*(z,\bt)
\end{aligned}
\end{equation}
where $\psi_0(z,\bt)=\diag(\exp(\xi(z,\bt_1)),\dots,\exp(\xi(z,\bt_N)))$,
and $\chi$ and $\chi^*$ are the bare Baker functions
with the following asymptotic expansion
\begin{equation}
\begin{aligned}
\label{asym}
\chi(z)&\sim 1+\beta z^{-1}+{\cal O}(z^{-2}),\quad
z\to\infty,\\
\chi^*(z)&\sim 1-\beta z^{-1}+{\cal O}(z^{-2}),\quad
z\to\infty ,
\end{aligned}
\end{equation}
and the matrix $\beta$ is given by
\begin{equation} \label{beta-c}
\begin{aligned}
\beta_{ii}(\bt)  &:=-\frac{\D\ln \tau(\bt)}{\D u_i}, i=1,\dots,N, \\
\beta_{ij}(\bt) &:=\varepsilon_{ij}\frac{\tau_{{ij}}(\bt)}{\tau(\bt)},
\quad
i\neq j,\, i,j=1,\dots,N,
\end{aligned}
\end{equation}
with $u_k:=t_{k,1}$, $k=1,\dots,N$.

Thus, by setting $\bell\to\bell+\be_i$ and $\bell'\to\bell-\be_j$
in (\ref{bi1}) one obtains
\begin{equation}\label{bi2}
\int_{S^1}\dif z\;\psi(z,\bt)\psi^*(z,\bt^\prime)=0 \; .
\end{equation}
The symmetry operators $X_i$ and $X_i^*$ of (\ref{bilineal}) induce
a corresponding action, say $\gX_i$ and $\gX_i^*$, respectively, on
Baker functions:
\begin{align*}
  \gX_i(p)\psi(z,\bt) & =[\gT_i^-(p)\psi(z,\bt)]\Big(-\frac{p}{z}\Big)^{P_i}, \\
  \gX_i^*(p)\psi(z,\bt) & =[\gT_i^+(p)\psi(z,\bt)]\Big(-\frac{z}{p}\Big)^{P_i}.
\end{align*}
Here  $P_i$ stands for the matrix with elements
$(P_i)_{jk}=\delta_{ij}\delta_{ik}$. Notice that in order that
$\gT_i^{\pm}\psi(z,\bt)$ be convergent it is required that
$|p|>|z|$.

Both bilinear identities (\ref{bi1}) and (\ref{bi2}) are useful for
characterizing
the $N$-component KP hierarchy. In particular, (\ref{bi2}) is suitable for
formulating the Grassmannian approach to the hierarchy \cite{sw},
which in turn
is very convenient in the derivation of the linear system of equations
for the Baker functions.
Let us, for instance, outline this approach for the Baker function $\psi$.
To this end, we denote by $W$  the set of $N\times N$ matrix functions
$\varphi(z)$ such that:
\[
\int_{S^1}\dif z\, \varphi(z)\psi^*(z,\bt')=0,
\]
for all $\bt'$ in the definition domain of $\psi^*$. Under
appropriate conditions the set $W$ belongs to an
infinite-dimensional Grassmannian manifold \cite{sw}. From
(\ref{bi2}) it follows that $W$ is a left $M_N(\C)$-module, with
$M_N(\C)$ being the ring of $N\times N$ complex matrices.  We shall
use the standard notation $E_{ij}$ for the linear basis in
$M_N(\C)$, and in particular $P_i=E_{ii}$. As a consequence of
(\ref{bi2}) and the form of the asymptotic expansion of $\psi^*$ as
$z\to\infty$ one  has that for any  $\bt$:
\begin{equation}\label{sd}
W=\bigoplus_{n\geq 0}M_N(\C)
\cdot v_n(\bt),\quad v_n(z,\bt)=
\Big(\sum_{k=1}^N\frac{\partial}{\partial u_k}\Big)^n\psi(z,\bt).
\end{equation}
Notice that
\begin{equation}\label{vn}
v_n(z)\sim (z^n+{\cal O}(z^{n-1}))\psi_0(z),\quad z\to\infty.
\end{equation}
Thus, the linear system for the $N$-component KP hierarchy results
from the decompositions of the time derivatives of $\psi$ in terms
$v_n$, $n=0,\dots,\infty$. In particular, by decomposing
$P_i\D\psi/\D u_k$, $i\neq k$, one gets
\begin{equation}
\label{lin-p}
\frac{\D \bpsi_i}{\D u_k}=\beta_{ik}\bpsi_k,
\end{equation}
with
\[
\bpsi_i:=(\psi_{i1},\dots,\psi_{iN}).
\]

Proceeding in a similar way for the adjoint Baker function we
arrive to
\begin{equation}
 \label{linad}
\frac{\D \bpsi_j^*}{\D u_k}=\bpsi_k^*\beta_{kj}, \quad j\neq k,
\end{equation}
where
\[
\bpsi_i^*:=\begin{pmatrix}
  \psi_{1i}^* \\\vdots\\
\psi_{Ni}^*
\end{pmatrix}.
\]

The compatibility of either (\ref{lin-p}) or (\ref{linad}) leads to
the Darboux equations for the $\beta$'s:
\begin{equation}\label{dar}
\frac{\D\beta_{ij}}{\D u_k}=\beta_{ik}\beta_{kj},\quad \text{$i$, $j$ and $k$ different.}
\end{equation}

As for the bilinear identity (\ref{bi1}),
the evaluation of the residue at infinity of the integrand
 provides the Hirota representation of the $N$-component
KP hierarchy. In particular,
\begin{align}
&\tau\frac{\D^2\tau}{\D u_i\D u_j}-
\frac{\D\tau}{\D u_i}\frac{\D\tau}{\D
u_j}-\tau_{{ij}}\tau_{{ji}}=0,\quad i\ne j,
\label{hir-c} \\
\label{tau-3}
&\tau\frac{\D\tau_{{ij}}}{\D u_k}-\tau_{ij}\frac{\D\tau}{\D
u_k}-\varepsilon_{ij}
\varepsilon_{ik}\varepsilon_{kj}\tau_{ik}\tau_{kj}=0 ,\quad
\text{$i,j$ and $k$ different},
\end{align}
being (\ref{tau-3}) the Hirota form  of the above Darboux
equations.

By setting  $\bell\to\bell+\be_i$ and
$\bell'\to\bell+\be_k-\be_l-\be_j$ in (\ref{bi1}) we obtain
\begin{equation} \label{tau-4}
\varepsilon_{ij}\varepsilon_{kl}\tau \: \Sgot_{ik}(\tau_{jl}) +
\varepsilon_{il}\varepsilon_{jk}\tau_{ik}\tau_{jl} -
\varepsilon_{ik}\varepsilon_{jl}\tau_{il}\tau_{jk} =0
\; , \; \quad \text{$i,j,k$ and $l$ different}.
\end{equation}
This relation, which can be found in  \cite{kv},  is just a Fay
trisecant formula for theta functions on Riemann surfaces
\cite{maffei}.

\section{Conjugate nets and quadrilateral latices}
\label{sec:nets}

The Darboux equations (\ref{dar}) for the so called rotation
coefficients $\beta_{ij}$ characterize $N$-dimensional
submanifolds of $\R^M$, $N\leq M$, parametrized by conjugate
coordinate systems (multiconjugate nets)~\cite{Darboux},
and are the compatibility
conditions of the following linear system
\begin{equation} \label{X}
\frac{\partial \bX_j}{\D u_i}=\beta_{ji} \bX_i,
\quad i,j=1,\dotsc,N,\quad i\ne j,
\end{equation}
involving  $M$-dimensional vectors $\bX_i$,
tangent to the coordinate lines. The so called Lam\'e coefficients
$H_i$ satisfy
\begin{equation*} \label{H}
\frac{\partial H_j}{\D u_i} = \beta_{ij} H_i, \quad i,j=1,\dotsc,N,
\quad i\ne j,
\end{equation*}
in terms of which  the points $\bx$ of the net  are found by
integrating the following equation
\begin{equation*} \label{points}
\frac{\D \bx}{\D u_i}= \bX_i H_i,\quad i=1,\dotsc, N.
\end{equation*}

Thus, given the Baker function $\psi$, one can construct
conjugate nets with $\beta$'s as appearing in (\ref{asym}) and with the
tangent vectors $\bX_i$ being the rows of the
matrix
\begin{equation} \label{X-psi}
X(\bt)=\int_{S^1}\dif
z\;\psi(z,\bt)f(z),
\end{equation}
for some distribution  matrix $f(z)\in M_{N\times M}(\C)$. Given
the adjoint Baker function $\psi^*$, the Lam\'e coefficients are
given by the entries of the row matrix
\begin{equation} \label{H-psi}
H(\bt)=\int_{S^1}\dif z\;g(z)\psi^*(z,\bt),
\quad i=1,\cdots, N,
\end{equation}
for some distribution row matrix $g(z)\in\C^N$.

Therefore we arrive to the following
\begin{pro} \label{cor:mKP-nets}
 The solutions of the N-component KP hierarchy describe $N$
dimensional conjugate nets with coordinates $u_i = t_{i,1}$,
$i=1,\dots,N$, while the remaning times $t_{i,k}$, for $k>1$,
describe integrable iso-conjugate deformations of the nets.
\end{pro}

In particular, for $N=2$, the Davey-Stewartson hierarchy describes
the iso-conjugate deformations of two dimensional conjugate
nets~\cite{Kon-DS}.

Transformations of conjugate nets have been extensively studied
in the literature \cite{Eisenhart} and the most convenient way to characterize
them is through the notion of congruences of lines.
The basic transformations of conjugate nets are listed below.

(i) The Laplace transformation $\cL_{ij}(\bx)$, $i\ne j$, of a
conjugate net $\bx$ is the $j$-th focal net of the $i$- th tangent
congruence of $\bx$ \cite{Darboux}; in simple terms it means that
the line tangent to the $i$-th coordinate line at a point $\bx$ of
the net is tangent to the $j$-th coordinate line of the transformed
net at the corresponding point $\cL_{ij}(\bx)$ (see Figure 1).

It turns out \cite{Darboux,TQL} that the position points of the transformed
net are given by
\begin{equation*}
\cL_{ij}(\bx) = \bx - \frac{H_j}{\beta_{ij}}\bX_i.
\end{equation*}
The corresponding transformation for
the rotation coefficients $\beta_{ij}$ are \cite{Ferapontov}
\begin{align}
\label{Lijij}\cL_{ij}(\beta_{ij})&=\beta_{ij} \left( \beta_{ij}\beta_{ji}-\frac{\D^2\log\beta_{ij}}{\D u_i\D u_j} \right), \\ 
\label{Lijji}\cL_{ij}(\beta_{ji})&=\frac{1}{\beta_{ij}}\label{t1},\\
\label{Lijki}\cL_{ij}(\beta_{ki})&=\frac{\beta_{kj}}{\beta_{ij}},\\ 
\label{Lijjk}\cL_{ij}(\beta_{jk})&=-\frac{\beta_{ik}}{\beta_{ij}},\\
\label{Lijik}\cL_{ij}(\beta_{ik})&=-\beta_{ij} \frac{\D}{\D u_i}\left(\frac{\beta_{ik}}{\beta_{ij}}\right),\\
\label{Lijkj}\cL_{ij}(\beta_{kj})&=\beta_{ij} \frac{\D}{\D u_j}\left(\frac{\beta_{kj}}{\beta_{ij}}\right),\\
\label{Lijkl}\cL_{ij}(\beta_{kl})&=\beta_{kl}-
\frac{\beta_{kj}\beta_{il}}{\beta_{ij}},
\end{align}
where all the indices, $i$, $j$, $k$ and $l$, are different. It can
be also shown that the Laplace transformations satisfy the
following relations \cite{Ferapontov}
\begin{align}
\cL_{ij}\circ \cL_{ji} & = \cL_{ji}\circ \cL_{ij} = \id , \label{lapl-seq} \\
\label{l-l}
\cL_{ij}\circ \cL_{jk} = \cL_{ik},  &\quad
\cL_{ij}\circ \cL_{ki} = \cL_{kj}.
\end{align}

We finally recall that the Laplace transformation $\cL_{ij}$ of a
2-dimensional conjugate net provides the geometric meaning~\cite{Darboux}
of the
2-dimensional Toda system. In fact, interpreting in
equations (\ref{t1}) and (\ref{lapl-seq})
the operator $\cL_{ij}$ as translation in
the discrete variable $n$, we obtain
\begin{equation*}
\frac{\D^2\log\beta_{ij}(n)}{\D u_i\D u_j} =
\frac{\beta_{ij}(n)}{\beta_{ij}(n-1)}
-\frac{\beta_{ij}(n+1)}{\beta_{ij}(n)}.
\end{equation*}

(ii) The L\'{e}vy transformation $\cL_i(\bx)$ of a conjugate net $\bx$
is a net conjugate to the $i$-th tangent congruence
\cite{Eisenhart}; i. e., the lines $\langle \bx , \cL_i(\bx)
\rangle $ are {\em tangent} to $i$-th coordinate lines at $\bx$
(see Figure 2).

The position points of the new net are given by \cite{Eisenhart,TQL}
\begin{equation} \cL_i(\bx) = \bx - \frac{\Omega[\zeta,H]}{\zeta_i}\bX_i,
\end{equation}
where $\zeta_k$, $k=1,\cdots,N$, is a solution of the linear system
(\ref{lin-p}), and ${\Omega[\zeta,H]}$ is a solution of the
equations
\begin{equation*}
\frac{\D\Omega}{\D u_k} = \zeta_k H_k, \quad k=1,\dots, N.
\end{equation*}
The corresponding transformations for the tangent vectors
$\bX_i$  are \cite{KoSchief}
\begin{equation}\label{TLevy}
\begin{aligned}
\cL_{i}(\bX_i) & = -\frac{\D\bX_i}{\D u_i} + \frac{1}{\zeta_i}
\frac{\D\zeta_i}{\D u_i} \bX_i,
 \\
\cL_{i}(\bX_k) & = \bX_k - \frac{\zeta_k}{\zeta_i}\bX_i, \quad k\ne i,\,
k=1,\dots,N.
\end{aligned}
\end{equation}
(iii) The adjoint L\'{e}vy transformation $\cL_i^*(\bx)$ of a conjugate
net $\bx$ is the $i$-th focal net of a congruence conjugate to
$\bx$ \cite{Eisenhart}; i. e., the lines $\langle \bx ,
\cL_i^*(\bx)\rangle $ are {\em tangent} to the $i$-th coordinate lines of the
new net. The position points of the new net are given by
\cite{Eisenhart,TQL}
\begin{equation*}
\cL_i^*(\bx) = \bx - \frac{\bOmega[\bX,\zeta^*]}{\zeta_i^*} H_i,
\end{equation*}
where $\zeta_k^*$, $k=1,\cdots,N$, is a solution to the adjoint
linear system (\ref{linad}) and ${\bOmega[\bX,\zeta^*]}$ is a
solution of the equations
\begin{equation*} \label{O-aL}
\frac{\D\bOmega}{\D u_k} = \bX_k\zeta^*_k, \quad k=1,\cdots, N.
\end{equation*}
The corresponding transformations for the tangent vectors
$\bX_i$  are \cite{KoSchief}
\begin{equation}\label{TadLevy}
\begin{aligned} 
\cL_{i}^*(\bX_i) & = - \frac{\bOmega[\bX,\zeta^*]}{\zeta_i^*}, \\
\cL_{i}^*(\bX_k) & =
\bX_k - \beta_{ki}\frac{\bOmega[\bX,\zeta^*]}{\zeta_i^*},\quad k\neq
i,\, k=1,\dots,N.
\end{aligned}
\end{equation}
(iv) The fundamental transformation $\cF(\bx)$ of a conjugate net
$\bx$ shares with $\bx$ the same conjugate congruence; i. e., the
lines $\langle \bx , \cF(\bx) \rangle $  {\em intersect} both nets
along the coordinate lines. It can be viewed as the composition of
L\'{e}vy and adjoint L\'{e}vy transformations $\cF_i = \cL_i\circ \cL_i^*$.
The fundamental transformation is given by \cite{Eisenhart}
\begin{equation*} \cF_i(\bx) =
\bx - \bOmega[\bX,\zeta^*]\frac{\Omega[\zeta,H]}{\Omega[\zeta,\zeta^*]};
\end{equation*}
here  $\Omega[\zeta,\zeta^*]$ is a solution of
\begin{equation*}
\frac{\D\Omega}{\D u_k} = \zeta_k \zeta^*_k, \quad k=1,\dots, N.
\end{equation*}
The corresponding transformations for the tangent vectors
$\bX_i$  are \cite{KoSchief}
\begin{equation}\label{fi}
\cF_i(\bX_j)=\bX_j-\frac{\bOmega[\bX,\zeta^*]
}{\Omega[\zeta,\zeta^*]}\zeta_j,\quad j=1,\dots,N.
\end{equation}
We first remark that L\'evy, adjoint L\'evy and Laplace transformations
are limiting cases of the fundamental transformation, in which one
of the two nets (or both nets) conjugate to the congruence of
the transformation are focal nets of the congruence~\cite{TQL}.

We also remark that, from Proposition~\ref{cor:mKP-nets}, these
transformations map solutions of the multicomponent KP hierarchy
into new solutions and, in this context, they were recently
investigated in~\cite{OevelSchief}, under the collective name of
``Darboux transformations", commonly used in the soliton
community~\cite{ms,KoSchief}.

It is a common belief that Darboux-type transformations of
integrable partial differential equations generate their natural
integrable discrete versions \cite{LeBen}. Furthermore, if the
original partial differential equation has a geometric meaning, the
Darboux-type transformations provide the natural discretization of
the corresponding geometric notions \cite{BoSchief,KoSchief2,TQL}.
For example, if we consider a conjugate net $\bx$ and two
fundamental transformations $\cF_1(\bx)$ and $\cF_2(\bx)$ of it,
the points
\[
\{ \bx, \cF_1(\bx), \cF_2(\bx), \cF_1(\cF_2(\bx))\}
\]
 are coplanar \cite{Eisenhart}.
It turns out that a lattice
\begin{alignat*}{2}
  \bx: &\,\Z^N &\rightarrow& \R^M\\
&\,\bn&\mapsto&\bx(\bn),
  \end{alignat*}
 $N\leq M$, whose elementary quadrilaterals are planar (i. e., a
quadrilateral lattice) is the correct discrete analogue of a
conjugate net \cite{Sauer,DQL1}. The planarity condition can be
expressed by the following linear equation (compare with (\ref{X}))
\begin{equation*} \label{X-D}
\Delta_j \bX_i=(T_jQ_{ij})\bX_j, \quad i,j=1,\dotsc,N,\;i\neq j,
\end{equation*}
being its compatibility conditions
\begin{equation}\label{disdar}
\Delta_kQ_{ij}=(T_kQ_{ik})Q_{kj},\quad \text{$i$, $j$ and $k$ different,}
\end{equation}
the discrete analogue of the Darboux equations (\ref{dar}). The
points $\bx$ of the lattice can be found by means of discrete
integration of
\[
\Delta_i\bx  = (T_iH_i) \bX_i, \quad i=1,\dots, N,
\]
where $H_i$ are solutions of equations
\[
\Delta_iH_j  = Q_{ij} T_iH_i, \; \; i,j=1,\dotsc,N,\;i\neq j.
\]
In the above formulas, $T_i$ is the translation operator in the
discrete variable $n_i$:
\[
 T_i f(n_1,\dots,n_i,\dots,n_N) =f(n_1,\dots,n_i+1,\dots,n_N),
 \]
and $\Delta_i = T_i - 1$ is the corresponding partial difference operator.

\section{Vertex operators as classical transformations of conjugate nets}

In this Section we show how the basic vertex operators associated
with the multicomponent KP hierarchy have a natural geometrical
interpretation as the classical transformations of  conjugate nets.

\subsection{
Identification between partial charge transformations in the zero
charge sector and Laplace transformations}

We first notice that the algebraic relations
(\ref{lapl-seq})-(\ref{l-l}) between the Laplace transformations
are the same as those satisfied by the $A_{N-1}$ root lattice
operators shifts (the Schlessinger transformations) $\Sgot_{ij}$
(\ref{ij-ji})-(\ref{ijk}). In fact, both transformations can be
identified as stated in the following:

\begin{pro} \label{prop:Laplace}
The root lattice shift $\Sgot_{ij}$ in the direction $\alpha_{ij}$
is the composition of the Laplace transformation $\cL_{ij}$ with a
trivial scaling symmetry of the Darboux equations.
\end{pro}

{\bf Proof:} Let us examine the Laplace transformation at the light
of $\tau$-functions,  by recalling the definition of the rotation
coefficients  (\ref{beta-c}). Starting from (\ref{Lijij}) we obtain
$\cL_{ij}\beta_{ij}=-\varepsilon_{ij}\beta_{ij}\D^2\log\tau_{ij}/\D
u_i\D u_j$, where we have used (\ref{hir-c}) in the form
$\tau^2\D^2\log\tau/\D u_i\D u_j=\tau_{ij}\tau_{ji}$. If in the
previous  identity we apply  $\Sgot_{ij}$ we get
$\tau_{ij}^2\D^2\log\tau_{ij}/\D u_i\D
u_j=(\Sgot_{ij}\tau_{ij})\tau$, so that
\[
\cL_{ij}\beta_{ij}=-\Sgot_{ij}\beta_{ij}.
\]

For the next three equations (\ref{Lijji})-(\ref{Lijjk}) the
following identifications trivially hold
\begin{align*}
\cL_{ij}\beta_{ji}&=-\Sgot_{ij}\beta_{ji},\\
\cL_{ij}\beta_{ki} &=
\varepsilon_{ki}\varepsilon_{kj}\varepsilon_{ij}\Sgot_{ij}\beta_{ki}, \\
  \cL_{ij}\beta_{jk} &=
\varepsilon_{ki}\varepsilon_{kj}\varepsilon_{ji}\Sgot_{ij}\beta_{jk}.
\end{align*}

The next two equations (\ref{Lijik}) and (\ref{Lijkj}) derive from
equation (\ref{tau-3}) once the shifts $\Sgot$ are applied
properly; namely, we have the identities:
\begin{align*}
\tau_{ik}\frac{\D\tau_{ij}}{\D u_i}-\tau_{ij}\frac{\D\tau_{ik}}{\D
u_i}&=\varepsilon_{kj}\varepsilon_{ki}\varepsilon_{ij}\tau\Sgot_{ik}\tau_{ij},\\
\tau_{kj}\frac{\D\tau_{ij}}{\D u_j}-\tau_{ij}\frac{\D\tau_{kj}}{\D
u_j}&=\varepsilon_{kj}\varepsilon_{ki}\varepsilon_{ij}\tau\Sgot_{kj}\tau_{ij},
\end{align*}
from where it follows
\begin{align*}
\cL_{ij}\beta_{ik}&=
\varepsilon_{ki}\varepsilon_{kj}\varepsilon_{ij}\Sgot_{ij}\beta_{ik},\\
\cL_{ij}\beta_{kj}&=
\varepsilon_{ki}\varepsilon_{kj}\varepsilon_{ji}\Sgot_{ij}\beta_{kj}.
\end{align*}

Finally,  (\ref{tau-4}) leads to
\[
\cL_{ij}\beta_{kl}=\varepsilon_{ki}\varepsilon_{kj}\varepsilon_{li}\varepsilon_{lj}\Sgot_{ij}\beta_{kl}.
\]

These results can be resumed as
\[
\frac{a_k}{a_l}\Sgot_{ij}\beta_{kl}=\cL_{ij}\beta_{kl},\quad
k,l=1,\dots,N,\;k\neq l,
\]
where:
\[
a_k=\varepsilon_{ki}\varepsilon_{kj}.
\]

At this point we must remark that the Darboux equations have the
following scaling symmetry $\beta_{ij}\to a_j/a_i\,\beta_{ij}$ that
comes from the freedom in the choice of the Lam\'{e} coefficients
$H_i\to a_i H_i$, where $a_i=a_i(u_i)$, $i=1,\dots,N$, are
functions of $u_i$ only.
\hfill$\Box$

Thus, a $\tau$-function of the multicomponent KP hierarchy describes
not only the integrable deformations of a single conjugate net but also
all its Laplace transforms.

%
%
%

\subsection{Vertex operators $\gX_i$ and $\gX_i^*$ as L\'{e}vy transformations}


Now we consider the action of the basic vertex operators at the
level of Baker functions.
 In the next proposition we identify the
action of the vertex operator $\gX_i(p)$ with the the classical
L\'evy transformation:

\begin{pro} \label{prop:Levy}
Given tangent vectors $\bX_j$, $i=j,\dots,M$, associated with the
Baker function $\psi(z,\bt)$ as prescribed in (\ref{X-psi}), then
\[
\cL_i(\bX_j)=p^{\delta_{ij}}\gT_i^-(p)(\bX_j),
\]
where $\cL_i$ stands for the L\'evy transformation with data
$\zeta_j:=\dfrac{\dif^n\psi_{ji}}{\dif z^n}(p)$, $j=1,\dots,M$,
where $n\geq 0$ is the order of the first non-zero $z$-derivative
of  the $i$-th column of $\psi$ at $p$.
\end{pro}
{\bf Proof:}
Firstly, we  observe that  the following asymptotic expansion
holds:
\begin{align}
\gT_i^-(p)\psi(z,t)&=\Big[\gT_i^-(p)\chi(z,\bt)\Big]\Big(1-
\frac{z}{p}P_i\Big)\psi_0(z,\bt)\label{p}\\
&=\Big[-\frac{z}{p}P_i+1-\frac{1}{p}\gT_i^-(\beta P_i)+ {\cal
O}\Big(\frac{1}{z}\Big)\Big]\psi_0(z,\bt),\notag
\end{align}
which can be compared with
\[
\frac{\D\psi}{\D u_i}=\Big[zP_i+\beta P_i+{\cal O}(\frac{1}{z}\Big)\Big]
\psi_0(z,\bt).
\]
Thus, as both $\gT_i^-(p)\psi$ and $\D\psi/\D u_i$ belong to the Grassmannian
element $W$ and taking into account (\ref{sd}) and (\ref{vn}) we deduce
\begin{equation}\label{t11}
\gT_i^-(p)\psi=-\frac{1}{p}\frac{\D\psi}{\D u_i}+\Big(1-\frac{1}{p}
(\gT_i^-(p)-1)\beta P_i\Big)\psi.
\end{equation}
Moreover, by using  the matrix form of equations (\ref{dar})
\[
P_j\frac{\D\psi}{\D u_i}=P_j\beta P_i\psi,\quad i\neq j,
\]
one gets
\[
\frac{\D\psi}{\D u_i}=P_i\frac{\D\psi}{\D u_i}+\sum_{j\neq i} P_j\beta P_i\psi=
P_i\frac{\D\psi}{\D u_i}+(\beta P_i-P_i\beta P_i)\psi.
\]
Hence (\ref{t11}) becomes
\begin{equation*}\label{t2}
\gT_i^-(p)\psi=\psi-\frac{1}{p}P_i\frac{\D\psi}{\D u_i}+\frac{1}{p}P_i\beta P_i\psi-\frac{1}{p}(\gT_i^-(p)\beta P_i)\psi,
\end{equation*}
or equivalently
\begin{equation}\label{t21}
P_j\gT_i^-(p)\psi=P_j\psi-\frac{1}{p}(\gT_i^-(p)\beta_{ji}E_{ji})\psi,\quad i\neq j,
\end{equation}
and
\begin{equation}\label{t22}
P_i\gT_i^-(p)\psi=-\frac{1}{p}P_i\frac{\D\psi}{\D u_i}+f(\bt)P_i\psi,
\end{equation}
where
\[
f(\bt):= 1+\frac{1}{p}(1-\gT_i^-(p))\beta_{ii}.
\]

From (\ref{p})  we notice that $\gT_i^-(p)\psi(p,\bt)P_i=0$, so
that by setting $z=p$ in (\ref{t21}) and (\ref{t22}) 
we conclude
\begin{align*}
\frac{1}{p}(\gT_i^-(p)\beta_{ji})(\bt)&=
\frac{\zeta_{j}(\bt)}{\zeta_i(\bt)},\label{v-xi}\\ f(\bt)&=\frac{1}{p}
\frac{\D\log\zeta_i}{\D u_i}(p,\bt),
\end{align*}
where
\begin{equation}\label{xibaker}
   \zeta_j(\bt):=\dfrac{\dif ^n\psi_{ji}}{\dif z^n}(p,\bt).
\end{equation}

 Therefore, we may rewrite (\ref{t21}) and (\ref{t22}) as
\begin{equation} \label{t3}
\begin{aligned}
(\gT_i^-(p)\bpsi_j)(z,\bt)&:=\bpsi_j(z,\bt)-
\frac{\zeta_j(\bt)}{\zeta_i(\bt)}\bpsi_i(z,\bt),
\quad j\neq i  \\
(\gT_i^-(p)\bpsi_i)(z,\bt)&:=-\frac{1}{p}\frac{\D\bpsi_i}{\D
u_i}(z,\bt)+\frac{1}{p}
\frac{\D\log\zeta_i}{\D u_i}(\bt)\bpsi_i(z,\bt).
\end{aligned}
\end{equation}
The rows of $\psi$ and $\gT_i^-(p)\psi$ provide tangent vectors for
conjugate nets, so that by comparing (\ref{t3}) with (\ref{TLevy})
we obtain
\[
\cL_i(\bpsi_j)=p^{\delta_{ij}}\gT_i^-(\bpsi_j).
\]
Hence, from (\ref{X-psi}) we get the desired result.
\hfill $\Box$

Now we identify the vertex operator $\gX^*_i(p)$ with the adjoint
L\'{e}vy transformation

\begin{pro} \label{prop:adLevy}
Given tangent vectors $\bX_j$, $j=1,\dots,M$, associated with the
Baker function as prescribed in (\ref{X-psi}), then
\[
\cL_i^*(\bX_j)=\frac{1}{p^{\delta_{ij}}}\gT_i^+(\bX_j),
\]
where $\cL_i^*$ stands for the adjoint L\'evy transformation with
data
\[
\zeta_j^*(\bt)=\frac{\dif^m\psi_{ij}^*}{\dif z^m}(p,\bt),\quad
j=1,\dots,M,
\]
with $m$ being the order of the first non-vanishing $z$-derivative
of the $i$-th row of $\psi^*$ at $z=p$, and potential
\[
\bOmega=-\frac{1}{p}[\gT_i^+(p)\bpsi_i]\zeta_i^*.
\]
\end{pro}
{\bf Proof:} On the one hand,  setting $\bt'=\bt-[1/p]\be_i$ in the
bilinear identity (\ref{bi2}) one gets
\[
(1-P_i)\gT_i^-(p)\beta-\beta(1-P_i)+pP_i-p\chi(p,\bt)P_i\gT_i^-(p)\chi^*(p,\bt)=0,
\]
which implies
\begin{gather*}
\beta_{ji}(\bt)=p\big[\gT_i^+(p)\chi_{ji}(p,\bt)\big]\chi_{ii}^*(p,\bt),\\
\beta_{ij}(\bt)=-p\big[\gT_i^-(p)\chi_{ij}^*(p,\bt)\big]\chi_{ii}(p,\bt),\\
\chi^*_{ii}(p,\bt)\gT_i^+(p)\chi_{ii}(p,\bt)=1.
\end{gather*}
for $j=1,\dots,N$ and $i\neq j$; hence,
\begin{gather*}
\begin{aligned}
\frac{[\gT_{i}^+(p)\psi_{ji}](p,\bt)}{[\gT_{i}^+(p)\psi_{ii}](p,\bt)}&=
\lim_{z\to p}\frac{[\gT_{i}^+(p)\psi_{ji}](z,\bt)}{
[\gT_{i}^+(p)\psi_{ii}](z,\bt)}=
\frac{[\gT_{i}^+(p)\chi_{ji}](p,\bt)}
{[\gT_{i}^+(p)\chi_{ii}](p,\bt)}=
\frac{\beta_{ji}(\bt)\chi_{ii}^*(p,\bt)}{p\chi^*_{ii}(p,\bt)}\\&=\frac{1}{p}
\beta_{ji}(\bt),
\end{aligned}\\
\begin{aligned}
\frac{[\gT_{i}^-(p)\psi_{ij}^*](p,\bt)}{[\gT_{i}^-(p)\psi_{ii}^*](p,\bt)}&=
\lim_{z\to p}\frac{[\gT_{i}^-(p)\psi_{ij}^*](z,\bt)}{[\gT_{i}^-(p)\psi_{ii}^*](z,\bt)}=
\frac{[\gT_{i}^-(p)\chi_{ij}^*](p,\bt)}{[\gT_{i}^-(p)\chi_{ii}^*](p,\bt)}=
-\frac{\beta_{ij}(\bt)\chi_{ii}(p,\bt)}{p\chi_{ii}(p,\bt)}\\
&=-\frac{1}{p}\beta_{ij}(\bt),
\end{aligned}\\
\frac{\D\log\gT_{i}^+(p)\psi_{ii}}{\D u_i}(p,\bt)=
\lim_{z\to p}\frac{\D\log\gT_{i}^+(p)\psi_{ii}}{\D u_i}(z,\bt)
=-\frac{\D\log\psi_{ii}^*}{\D u_i}(p,\bt),
\end{gather*}
for $j=1,\dots,N$ and $i\neq j$. By using l'H\^{o}pital rule with
$\zeta_j(\bt):=\dfrac{\dif^n\psi_{ji}}{\dif z^n}(p,\bt)$ and
$\zeta_j^*(\bt):=\dfrac{\dif^n\psi_{ij}^*}{\dif z^m}(p,\bt)$, where
$n$ and $m$ are the orders of first non-vanishing derivatives of
the $z$-derivatives of the $i$-th column of $\psi$ and $i$-th row
of $\psi^*$, respectively, we get the identities:
\begin{gather}\label{rela}
\frac{\gT_{i}^+(p)\zeta_j}{\gT_{i}^+(p)\zeta_i}=\frac{1}{p}
\beta_{ji},\notag\\
\frac{\gT_{i}^-(p)\zeta_j^*}{\gT_{i}^-(p)\zeta_i^*}=
-\frac{1}{p}\beta_{ij},\\ \notag
\frac{\D\log\gT_{i}^+(p)\zeta_i}{\D u_i}=-\frac{\D\log\zeta_i^*}{\D u_i}
\end{gather}
with $j=1,\dots,N$ and $i\neq j$.

On the other hand, from (\ref{t3}) it follows that
\begin{equation*}
\begin{aligned}
\bpsi_j  &= \gT_{i}^+(p)\bpsi_j -\frac{\gT_{i}^+(p)\zeta_j}{
\gT_{i}^+(p)\zeta_i}\gT_{i}^+(p)\bpsi_i,\quad j\neq i,  \\
\bpsi_i  &=-\frac{1}{p}\frac{\D\gT_{i}^+(p)\bpsi_i}{\D u_i}+\frac{1}{p}
\frac{\D\log\gT_{i}^+(p)\zeta_i}{\D u_i}\gT_{i}^+(p)\bpsi_i.
\end{aligned}
\end{equation*}

Therefore, these relations  become
\begin{equation}\label{relaciones}
\begin{aligned}
\bpsi_j  &= \gT_{i}^+(p)\bpsi_j -\frac{1}{p}\beta_{ji}\gT_{i}^+(p)
\bpsi_i,\quad j\neq i,\\
\bpsi_i  &=-\frac{1}{p}
\frac{1}{\zeta^*_i}\frac{\D\zeta^*_i\gT_{i}^+(p)\bpsi_i}{\D u_i}.
\end{aligned}
\end{equation}
In order to identify them with the L\'{e}vy transformations it is
required to introduce the potential
\[
\bOmega:=-\frac{1}{p}[\gT_{i}^+(p)\bpsi_i]\zeta^*_i.
\]
 The second equation in (\ref{relaciones}) can be written as $\dfrac{\D\bOmega}{\D
u_i}=\bpsi_i\zeta^*_i$; now, we proceed to identify the other
partial derivatives:
\begin{align*}
 -p\frac{\D\bOmega}{\D u_j}&=
=[\gT_{i}^+(p)\beta_{ij}][\gT_{i}^+(p)\bpsi_j]\zeta^*_i+
\beta_{ji}[\gT_{i}^+(p)\bpsi_i]\zeta^*_j\\
&=-p[\gT_{i}^+(p)\bpsi_j]\zeta^*_j-p\big(\bpsi_j-\gT_{i}^+(p)\bpsi_j)
\zeta_j^*\\ &=-p\bpsi_j\zeta^*_j,\quad j\neq i.
\end{align*}
where we have used equations (\ref{rela}) and (\ref{relaciones}).
Thus, we deduce that $\bOmega$ is characterized up to a constant
vector by
\[
\frac{\D\bOmega}{\D u_j}=\bpsi_j\zeta^*_j,\quad j=1,\dots,N.
\]

With the aid of $\bOmega$ we express (\ref{relaciones}) as
\begin{equation*}
\begin{aligned}
\gT_{i}^+(p)\bpsi_j &=\bpsi_j-\beta_{ji}\frac{\bOmega}{\zeta^*_i},
\quad j\neq i, \\
\gT_{i}^+(p)\bpsi_i &=- p \frac{\bOmega}{\zeta^*_i}.
\end{aligned}
\end{equation*}
Therefore, by comparing with (\ref{TadLevy}) we have
\[
\cL_i^*(\bpsi_j)=\frac{1}{p^{\delta_{ij}}}\gT_i^+(p)(\bpsi_j),
\]
where $\cL_i^*$ stands for the adjoint L\'{e}vy transformation with
data $\zeta_j^*$.\hfill $\Box$

\paragraph{Remarks:}
\begin{enumerate}
\item For the one component KP hierarchy our results for the
L\'{e}vy transformations reduces to those in \cite{am} for the Darboux
transformations.
\item
Notice that we might consider integer powers of the vertex
operators $\gT_i^-(p)$, say $\gT_i^-(p)^{n_i}$, that models the
shift $f(\bt)\to f(\bt-n_i[1/p]\be_i)$. From our Proposition
\ref{prop:Levy} is clear that it can be thought as a sequence of
L\'{e}vy transformations, that we will use in \S 5. However, we stress
that, even when the initial Baker function $\psi$ does not vanish
at $p$, its transformed function
 does; hence, we should take its $z$-derivative  at $p$
 to get the new transformation data. Thus, the sequence is defined in
terms of the truncated jet of the initial Baker function:
\[
\Big\{\psi_{ji}(p,\bt),\dfrac{\dif\psi_{ji}}{\dif
z}(p,\bt),\dots,\dfrac{\dif^{n_i-1}\psi_{ji}}{\dif
z^{n_i-1}}(p,\bt)\Big\}.
\]
\item In matrix terms the above propositions can be resumed as
\begin{align*}
  \cL_i(\psi) &= p^{P_i}[\gX_i(p)\psi]\Big(-\frac{z}{p}\Big)^{P_i},\\
\cL_i^*(\psi) &= \Big(\frac{1}{p}\Big)^{P_i}[\gX_i^*(p)\psi]
\Big(-\frac{p}{z}\Big)^{P_i}.
\end{align*}
\item We  notice that the correspondences provided by the last
two propositions,  derived from the bilinear equation for Baker
functions, are direct consequences, when $i\neq j$, of Fay
identities for the $\tau$-function. Namely, for the L\'{e}vy
transformation the relevant Fay identity is
\[
\begin{aligned}
&\varepsilon_{jk}z^{\delta_{jk}-1}
p^{\delta_{ki}}
\tau_{jk}(\bt-[1/z]\be_k)
\tau(\bt-[1/p]\be_i)\\
&+\varepsilon_{ki}\varepsilon_{ji}z^{\delta_{ik}-1}
p^{\delta_{ki}+\delta_{ij}-1}
\tau_{ik}(\bt-[1/z]\be_k)\tau_{ji}(\bt-[1/p]\be_i)\\
&-\varepsilon_{jk}z^{\delta_{jk}-1}
(p-z)^{\delta_{ki}}
\tau_{jk}(\bt-[1/z]\be_k-[1/p]\be_i)\tau(\bt)=0,\quad i\neq j.
\end{aligned}
\]
For the adjoint L\'{e}vy transformation the corresponding Fay identity
is
\[
\begin{aligned}
&\varepsilon_{jk}z^{\delta_{jk}-1}
(p-z)^{\delta_{ki}}\tau_{jk}(\bt-[1/z]\be_k)\tau(\bt+[1/p]\be_i)\\
&-\varepsilon_{ki}\varepsilon_{ji}z^{\delta_{ik}-1}
p^{\delta_{ki}-1}
\tau_{ik}(\bt-[1/z]\be_k+[1/p]\be_i)\tau_{ji}(\bt)\\
&-\varepsilon_{jk}z^{\delta_{jk}-1}p^{\delta_{ik}}
\tau_{jk}(\bt-[1/z]\be_k+[1/p]\be_i)\tau(\bt)=0,\quad i\neq j.
\end{aligned}
\]

\end{enumerate}

\subsection{The soliton vertex operator as fundamental transformation}

In the context of the  $\tau$-function theory, the soliton
solutions are generated by composite vertex operators which  are
infinitesimally generated by $b_i(p)c_j(q)$. It what follows we
will concentrate on  the diagonal case $i=j$:
\[
X_i(p,q)\tau(\bell,\bt):=\Big(\frac{p}{q}\Big)^{\ell_i}
\exp(\xi(p,\bt_i)-\xi(q,\bt_i))\tau(\bell,\bt-[1/p]\be_i+[1/q]\be_i).
\]
Since $X_i^2=0$ the exponential action reduces to
\[
\exp(aX_i(p,q))=1+aX_i(p,q).
\]
We shall show here that it correspond to a  fundamental
transformation. To this end we need the following
\begin{pro}
The $\tau$-function satisfies the following  identities for any
$i,j,k\in 1,\dots,N$:
\begin{multline}\label{fay}
\begin{aligned}
&\varepsilon_{jk}z^{\delta_{jk}-1}\frac{\tau_{jk}(\bt-[1/z]\be_k)
}{\tau(\bt)}\\&
-\varepsilon_{ik}\varepsilon_{ji}z^{\delta_{ik}-1}p^{\delta_{ji}-1}
\frac{1-\dfrac{p}{q}}{\Big(1-\dfrac{z}{q}\Big)^{\delta_{ki}}}
\frac{\tau_{ik}(\bt-[1/z]\be_k+[1/q]\be_i)}{
\tau(\bt-[1/p]\be_i+[1/q]\be_i)}
\frac{\tau_{ji}(\bt-[1/p]\be_i)}{\tau(\bt)}
\end{aligned}
\\
=\varepsilon_{jk}z^{\delta_{jk}-1}\Big(\frac{p}{q}\Big)^{\delta_{ij}}
\left(\frac{1-\dfrac{z}{p}}{1-\dfrac{z}{q}}\right)^{\delta_{ki}}
\frac{\tau_{jk}(\bt-[1/z]\be_k-[1/p]\be_i+[1/q]\be_i)}{\tau(\bt-[1/p]\be_i
+[1/q]\be_i)},
\end{multline}
\end{pro}
{\bf Proof:} On the one hand, as one can readily check from
(\ref{baker1}), the right hand side of (\ref{fay}) is, up to
exponential factors, just the $\tau$-function representation of the
components of the following vector
\[
\Big(\frac{p}{q}\Big)^{\delta_{ij}}\gT_i^-(p)\gT_i^+(q)\bpsi_j.
\]
On the other hand, we know that this is the composition of an
adjoint L\'{e}vy transformation, with transforming function
$\zeta^*_j(\bt)=\psi_{ij}^*(q,\bt)$ and potential $\bOmega(z)=-
\dfrac{1}{q}[\gT_i^+(q)\bpsi_i(z)]\zeta^*_i$, and a L\'{e}vy
transformation with   data
$\zeta_j(\bt)=(1-p/q)^{\delta_{ij}}\psi_{ji}(p,\bt)$ (the Baker
function after the first adjoint L\'{e}vy transformation is obtained
from $\gT^+_i(q)\psi$ by suitable normalization). Such
a composition 
 is a fundamental transformation:
\[
\cF_i(\bpsi_j)=\bpsi_j-\frac{\bOmega}{\Omega}\zeta_j,\quad
j=1,\dots, N,
\]
where
\
\begin{align*}
  \bOmega & =-\dfrac{1}{q}[\gT_i^+(q)\bpsi_i]\zeta_i^*, \\
  \Omega & =-\dfrac{1}{q}[\gT_i^+(q)\zeta_i]\zeta_i^*.
\end{align*}
That is to say, the matrix elements of the Baker function satisfy
\[
\psi_{jk}(z)-\frac{\gT_i^+(q)\psi_{ik}(z)}{\gT_i^+(q)\psi_{ii}(p)}
\psi_{ji}(p)=
\Big(\frac{p}{q}\Big)^{\delta_{ij}}\gT_i^-(p)\gT_i^+(q)\psi_{jk}(z),
\quad i,j,k=1,\dots, N.
\]
After substituting the expression (\ref{baker1}) of the Baker
function in terms of the $\tau$-function we obtain the desired
identities.\hfill $\Box$

From (\ref{fay})   one arrives to the following result
\begin{pro}
Given tangent vectors $\bX_j$, $i=j,\dots,M$, associated with the
Baker function $\psi(z,\bt)$ as prescribed in (\ref{X-psi}), the
induced action of the  operator $\exp(aX_i(p,q))$ is given by the
fundamental transformation
\[
\cF_i(\bX_j)=\bX_j-\frac{\bOmega}{\Omega}\zeta_j,\quad j=1,\dots,N
\]
with transforming data $\zeta_j(\bt)=(1-p/q)\psi_{ji}(p,\bt)$ and
$\zeta^*_j(\bt)=\psi^*_{ij}(q,\bt)$, $j=1,\dots,N$, and potentials:
\begin{align*}
  \bOmega & =-\dfrac{1}{q}[\gT_i^+(q)\bX_i]\zeta^*_i, \\
  \Omega &= -\dfrac{1}{a}\dfrac{q^{\ell_i-1}}{p^{\ell_i}}-
  \dfrac{1}{q}[\gT_i^+(q)\zeta_i]\zeta^*_i.
\end{align*}
\end{pro}
{\bf Proof:} In the expression (\ref{baker1}) substitute the old
$\tau$-function by the new one $(1+aX_i(p,q))\tau$, paying
particular attention, in the numerator, to the action of
$\gT_k^-(z)$ on this new function. Then, using (\ref{fay}), the
definition (\ref{baker1}) and comparing with (\ref{fi}) we obtain
the desired result.\hfill $\Box$

\paragraph{Remarks}
\begin{enumerate}
\item In the last two propositions we are assuming that $p$ and
$q$ are generic points for the corresponding Baker functions; i. e.
$p$ and $q$ are not zeroes of $\psi$ and $\psi^*$, respectively.

\item
An alternative derivation of (\ref{fay}) follows from the bilinear
equation (\ref{bi1}) by choosing $\bt'$ and $\bell'$ appropriately
and evaluating the corresponding residues of the integrand. In fact
they constitute a typical set of Fay identities:
\[
\begin{aligned}
&\varepsilon_{jk}z^{\delta_{jk}-1}p^{\delta_{ki}-1}q^{\delta_{ij}-1}
(z-q)^{\delta_{ki}}\tau_{jk}(\bt-[1/z]\be_k)\tau(\bt-[1/p]\be_i+[1/q]\be_i)\\
&-\varepsilon_{ki}\varepsilon_{ji}z^{\delta_{ik}-1}
p^{\delta_{ki}-1}q^{\delta_{ki}+\delta_{ij}-2}(p-q)
\tau_{ik}(\bt-[1/z]\be_k+[1/q]\be_i)\tau_{ji}(\bt-[1/p]\be_i)\\
&-\varepsilon_{jk}z^{\delta_{jk}-1}p^{\delta_{ij}-1}q^{\delta_{ki}-1}
(z-p)^{\delta_{ki}}\tau_{jk}(\bt-[1/z]\be_k-[1/p]\be_i+[1/q]\be_i)\tau(\bt)=0.
\end{aligned}
\]
\item Observe the presence of the parameter $a$ in the expression
of the potential $\Omega$. It plays the role of an integration
constant corresponding to the formula $\dfrac{\D\Omega}{\D
u_j}=\zeta_j\zeta_j^*$, $j=1,\dots, N$.
\item
 This result
strongly suggest a similar statement for the more general soliton
 operator: $\exp(1+a b_i(p)c_j(q))$, but here we should have
a composition of L\'{e}vy and adjoint L\'{e}vy in different directions and
the potentials would have now general integration constants.
\end{enumerate}

\section{Miwa transformation and  quadrilateral lattices}

Let us consider the bilinear identity (\ref{bi2}) for the Baker
function $\psi(z,\bt)$ and its adjoint $\psi^*(z,\bt)$. For each
complex number $p$ we can introduce new functions  depending on
 $N$ additional  discrete variables, $\bn\in\Z^N$, by defining
\[
\Psi(z,\bt,\bn):=\psi(z,\bt-\bn[1/p]),\quad \Psi^*(z,\bt,\bn):=\psi^*(z,\bt-\bn[1/p])
\]
where we understand that
\[
\bt-\bn[1/p]=(\bt_1-n_1[1/p],\bt_2-n_2[1/p],\dots,\bt_N-n_N[1/p]).
\]
Obviously (\ref{bi2}) becomes a continuous-discrete bilinear
equation of the form
\begin{equation*}\label{bild}
\int_{S^1}\dif z\;\;\Psi(z,\bt,\bn)\Psi^*(z,\bt',\bn^\prime)=0,
\end{equation*}
for any $\bt,\bt'\in\Cninf$ and $\bn,\bn'\in\Z^N$.

From (\ref{baker}) it follows that
\begin{equation*}
\begin{aligned}
\Psi(z,\bt,\bn)&:=\Xi(z,\bt,\bn)\Psi_0(z,\bt,\bn),\\
\Psi^*(z,\bt,\bn)&:=\Psi_0(z,\bt,\bn)^{-1}\Xi^*(z,\bt,\bn)
\end{aligned}
\end{equation*}
where
\[
\Psi_0(z,\bt,\bn):=\psi_0(z,\bt)
\diag\bigg(\Big(1-\frac{z}{p}\Big)^{n_1},
\dots,\Big(1-\frac{z}{p}\Big)^{n_N}\bigg),
\]
and
\begin{align*}
 \Xi(z,\bt,\bn)&:=\chi(z,\bt-\bn[1/p]),\\
\Xi^*(z,\bt,\bn)&:=\chi^*(z,\bt-\bn[1/p]),
\end{align*}
 have the following
asymptotic expansion
\begin{equation*}
\begin{aligned}
\Xi(z)&\sim 1- p Q z^{-1}+{\cal O}(z^{-2}),\quad
z\to\infty,\\
\Xi^*(z)&\sim 1+p Q z^{-1}+{\cal O}(z^{-2}),\quad
z\to\infty ,
\end{aligned}
\end{equation*}
with
\[
Q_{ij}(\bt,\bn):=-\frac{1}{p}\beta_{ij}(\bt-\bn[1/p]).
\]
If we fix our attention on the $\bn$ dependence the asympotic
module structure is now
\[
W=\bigoplus_{n\geq 0} M_N(\C)\cdot v_n(\bn), \quad
v_n(z,\bn)=\Big(\sum_{k=1}^N \Delta_k\Big)^n\Psi(z,\bn).
\]
The linear systems for $\Psi$ follow from the decomposition of the
discrete derivatives of $\Psi$ in terms of $v_n$.  A similar
analysis holds for $\Psi^*$ and we obtain
\begin{pro}
The objects $Q$, $\Psi$ and $\Psi^*$ do satisfy
\begin{align} \label{lindis}
\Delta_k \bPsi_{i}&=(T_k Q_{ik})\bPsi_{k},\quad i\ne k,\\
\Delta_k \bPsi_{j}^*&= Q_{kj}(T_k\bPsi_{k}^*),\quad j\ne k.
\end{align}
\end{pro}

\paragraph{Remarks}
\begin{enumerate}
\item
Observe that (\ref{lindis}) has been already proved in the first
formula of (\ref{relaciones}), where one should apply $\gT_i^-(p)$
and perform the replacement $\gT_i^-(p)\to T_i$.

\item The compatibility of (\ref{lindis}) gives the discrete Darboux
equation (\ref{disdar}). It is clear that $\bX_i(\bn)$ and
$H_i(\bn)$ can be obtained by the analogues of equations
(\ref{X-psi}) and (\ref{H-psi}), respectively. Hence,
 we have a quadrilateral lattice in the discrete
 variable $\bn$. From a geometrical point of view this has a clear
interpretation.

As we mentioned in Section 3, the Darboux-type transformations of
soliton equations provide their integrable discretization
\cite{LeBen}.  In the present Miwa-like scheme the translation
$T_i$ in the $n_i$ variable corresponds to the vertex operator
$\gT_i^-(p)$. Since, from Proposition~\ref{prop:Levy}, $\gT_i^-(p)$
corresponds to a Levy transformation then $\bx(\bn)$ describes a
quadrilateral lattice (see Figure 3).

\item
Obviously our approach gives, through the Miwa transformation, a
$\tau$-function formulation of the quadrilateral lattices and  a
quantum field theoretical representation of them in terms of
$b$-$c$ systems. For completeness, we give the $\tau$-function
expression of the quadrilateral lattice equation (\ref{disdar}):
\begin{align*}
&(T_i\tau)(T_j\tau)-
\tau(T_iT_j\tau)-(T_i\tau_{{ij}})(T_j\tau_{{ji}})=0, \quad i\ne j,
\label{hir-d} \\
\label{tau-3d}
&\tau(T_k\tau_{{ij}}) -(T_k\tau)\tau_{{ij}} -\varepsilon_{ij}
\varepsilon_{ik}\varepsilon_{kj}(T_k\tau_{{ik}})\tau_{{kj}}=0 ,\quad
\text{$i,j$ and $k$ different}.
\end{align*}
\end{enumerate}

\section*{Acknowledgements}
A. D.  acknowledges partial support from KBN grant no. 2P03 B
18509. M. M., L. M. A. and E. M. acknowledges  partial support from
CICYT {\em proyecto} PB95--0401 and M. M. and L. M. A. from the
exchange agreement between {\em Universit\` a La Sapienza }  of
Rome and {\em Universidad Complutense} of Madrid.

\end{document}